\documentclass[review]{elsarticle}

\usepackage{lineno,hyperref}
\modulolinenumbers[5]

\journal{Journal of \LaTeX\ Templates}

\usepackage{bm}
\usepackage{array}
\usepackage{amsmath}
\usepackage{amssymb}
\usepackage{subfigure}

\usepackage{footnpag}			      	
\usepackage{float}
\usepackage{comment}

\begin{document}

\begin{frontmatter}

\title{Six Degree-of-Freedom Body-Fixed Hovering over Unmapped Asteroids via LIDAR Altimetry and Reinforcement Meta-Learning}
\tnotetext[mytitlenote]{Accepted for 2019 AAS/AIAA Astrodynamics Specialist Conference}

\author{Brian Gaudet\fnref{myfootnote1}}
\cortext[mycorrespondingauthor]{Corresponding author}
\ead{briangaudet@mac.com}
\fntext[myfootnote1]{Department of Systems and Industrial Engineering, University of Arizona}

\author{Richard Linares\fnref{myfootnote2}}
\fntext[myfootnote2]{Department of Aeronautics and Astronautics, Massachusetts Institute of Technology}

\author{Roberto Furfaro\fnref{myfootnote3}}
\fntext[myfootnote3]{Department of Systems and Industrial Engineering, Department of Aerospace and Mechanical Engineering, University of Arizona}

\begin{abstract}
We optimize a six degrees of freedom hovering policy using reinforcement meta-learning. The policy maps flash LIDAR measurements directly to on/off spacecraft body-frame thrust commands, allowing hovering at a fixed position and attitude in the asteroid body-fixed reference frame. Importantly, the policy does not require position and velocity estimates, and can operate in environments with unknown dynamics, and without an asteroid shape model or navigation aids. Indeed, during optimization the agent is confronted with a new randomly generated asteroid for each episode, insuring that it does not learn an asteroid's shape, texture, or environmental dynamics. This allows the deployed policy to generalize well to novel asteroid characteristics, which we demonstrate in our experiments. Moreover, our experiments show that the optimized policy  adapts  to actuator failure and sensor noise. Although the policy is optimized using randomly generated synthetic asteroids, it is tested on two shape models from actual asteroids: Bennu and Itokawa. We find that the policy generalizes well to these shape models. The hovering controller has the potential to simplify mission planning by allowing asteroid body-fixed hovering immediately upon the spacecraft's arrival to an asteroid. This in turn simplifies shape model generation and allows resource mapping via remote sensing immediately upon arrival at the target asteroid. 

\end{abstract}
\begin{keyword}
Reinforcement Learning\sep Asteroid Missions\sep Hovering 
Artificial Intelligence\sep Autonomous Maneuvers
\end{keyword}

\end{frontmatter}


\section{Introduction}
Recently there has been increased interest in robotic missions to near Earth asteroids, for both scientific and commercial purposes. The prevalent concept of operations requires complete characterization of the asteroid's shape and dynamics prior to a sample return maneuver. Before a shape model can even be generated, the environmental dynamics must be characterized to a high degree of accuracy in order to allow calculation of stable orbits from which shape model generation takes place \cite{schutz2004statistical}.  Moreover, these stable orbits are in general only possible over a limited range of latitudes \cite{lara2002stability}. Asteroid body-fixed hovering at arbitrary locations in proximity to the asteroid has the potential to simplify mission planning, allowing high resolution sensor measurements at arbitrary locations \cite{broschart2005control}. Hovering in the inertial frame with the asteroid rotating below the spacecraft is possible at arbitrary altitudes (within the limits of terrain hazards) and in general will require less fuel than hovering in the asteroid body-fixed frame.  Although hovering in the asteroid body-fixed frame requires more fuel expenditure and cannot be performed at large distances from the asteroid, body frame hovering has the advantage of allowing multiple sensor measurements from a fixed position with respect to the asteroid. Finally, body fixed hovering close to the surface would allow a spacecraft to drill or collect surface samples while compensating for the force induced by the manipulators. Clearly, both types of hovering would be useful for asteroid missions. 

Previous work in hovering in close proximity to asteroids includes [\citenum{sawai2002control}], where the authors develop a hovering controller that uses altimetry measurements to hover in the asteroid body-fixed frame. Their work uses a single altimeter and thrusting direction, but assumes the sensor is aligned with the gravitational acceleration at the hovering point, the altitude is below the resonance radius (the altitude where gravitational and centrifugal forces cancel), and that the centrifugal force components perpendicular to thrust direction are known. Furfaro develops a 3-DOF hovering controller using sliding mode control theory \cite{furfaro2014hovering}. In other work \cite{gaudet2014real} Gaudet and Furfaro demonstrate both hovering and TAG maneuvers using a Rao-Blackwellized particle filter to infer the spacecraft's position and velocity using altimetry measurements and an asteroid shape model. Lee et. al. demonstrates 6-DOF hovering using a control law developed in the Lie group SE(3) \cite{lee2014almost}, but their method requires an estimate of the environmental dynamics. Gaudet and Furfaro developed a 3-DOF hovering controller using reinforcement learning \cite{gaudet2012robust} that showed improved transient response as compared to an LQR controller.  Importantly, previous work does not cover the case where the spacecraft arrives at an asteroid and we want the spacecraft to be able to immediately hover in the body-fixed frame in the case where 1.) there is no knowledge of the environmental dynamics and 2.) there is not an existing shape model that can be used by a navigation system to infer the spacecraft's position and velocity.

Inertial hovering has been successfully executed in both Hayabusa missions. The most recent, Hayabusa 2 \cite{tsuda2013system} arrived at the asteroid Ryugu in June 2018 and, after a sequence of close proximity operations including asteroid mapping and surface's touchdown and sample collection, departed the celestial body in November 2019. It is expected to deliver the sample to Earth in late 2020. One of the major modes of operation included the ability of the spacecraft to hover at different altitudes for either surface mapping and/or in preparation for the touchdown sequence.  The spacecraft employs a combination of Reaction Wheels (RW) and RCS thrusters to control attitude and position. A wide angle camera called ONC-W \cite{kameda2017preflight} is employed for navigation purposes. The camera is coupled with a dedicated image processor. Navigation has two major modes: the Asteroid Image Tracking (AIT) mode which calculates the image center of the asteroid Ryugu when in the Field of View (FOV). Conversely, in the Target Marker Mode (TMT) mode, the ONC-W tracks a target marker previously deployed on the asteroid surface \cite{yasuda2020operational}. A LIDAR system is employed to measure the spacecraft altitude \cite{mizuno2017development} . The latter is generally used at distance larger than 50 meters. For lower altitudes (5-50 meters), a Laser Range Finder (LRF) is employed. In a home position of about 20 km, hovering is executed by  a Ground Control Point Navigation (GCP-NAV) which employs the AIT mode \cite{yasuda2020operational}. Indeed the ONC-W sends images to the ground every ten minutes. A ground operator manually overlay the asteroids estimated shape and GCPs to the image to estimate the spacecraft position. Subsequently, the spacecraft position is propagated forward to account for the communication time delay. Eventually, the required delta-V is uploaded to the spacecraft for timed execution. Once the spacecraft is hovering below 50 meters, the TMT mode is executed by a combination of ground and on-board operations. In this phase,the position of a pre-deployed surface marker (reflector) is autonomously computed on-board. At this stage, hovering is controlled in a 6-DOF fashion using attitude and navigation information. Hovering generally occurred above the marker. For the final descent, although the team  had originally planned to hover at 25 meters altitude, flight data showed that the hovering occurred at an altitude of 8.5 meters \cite{yoshikawa2020hayabusa2}.

In this work we focus on the body-fixed hovering problem where neither a shape model nor information about the environmental dynamics are available.  Without a shape model, which allows a navigation system to infer the spacecraft's position in the asteroid body-fixed frame, body-fixed hovering is a challenging problem that to our knowledge has not yet been solved. The chief difficulty is that as the asteroid rotates it induces hovering position errors, and the hovering policy must learn how to correct for these errors by observing the changing LIDAR altimetry readings and use its recollection of these changing sensor readings to correct the hovering position error.  The problem is further complicated by pulsed thrusters, which will likely cause an overshoot with corrective thrust commands. 

The goal is to remain at a constant asteroid body-fixed position and attitude from the commencement of the hovering maneuver. We will assume that the spacecraft is equipped with  a flash LIDAR  system, gyroscopes that can measure the change in the spacecraft's attitude from the initiation of the hovering maneuver, and rate gyros that measure rotational velocity.  We further assume that these sensors can provide measurements every 6s. At the start of the hovering maneuver, the spacecraft is pointed in the general direction of the asteroid, and consequently at least some of the flash LIDAR elements can return valid altimeter readings. A possible concept of operations would be for the spacecraft to slowly approach the asteroid using a navigation system that keeps the asteroid centered in a camera's field of view, and then commence hovering when the mean range of the flash LIDAR elements indicates an acceptable hovering altitude. What happens next is mission specific, potential low altitude scenarios include the spacecraft hovering close to the surface to release a beacon or rover, or collect samples.  Potential high altitude hovering scenarios include shape model generation (where the ability to take multiple readings from the same position should simplify simultaneous location and mapping), remote sensing, as well as tagging the landing site with a targeting laser to facilitate a precision landing by a separate lander, as described in \cite{gaudet2019seeker}.

Our hovering controller is optimized using reinforcement learning (RL), which learns a policy that maps sensor measurements directly to on/off thrust commands, and that can adapt both to unknown environmental dynamics and novel asteroid shapes and textures. The policy is learned through simulated interaction between an environment and an agent instantiating the policy.  Adaptability is achieved through RL-Meta Learning (Meta-RL) \cite{mishra2017simple,frans2017meta,wang2016learning}, where different asteroid shapes and environmental dynamics are treated as a range of partially observable Markov decision processes (POMDP). In each POMDP, the policy's recurrent network hidden state evolves over the course of an episode based off of the history of observations and actions, capturing information about hidden variables that are useful in minimizing the cost function; these include asteroid shape, texture, environmental dynamics, and changes in the spacecraft's internal dynamics.  By optimizing the policy over this range of POMDPs, the trained policy will be able to adapt in real time to novel POMDPs encountered during deployment. Specifically, even though the policy's parameters are fixed after optimization, the policy's hidden state will evolve based off the history of observations and actions experienced in the current POMDP, thus adapting to the environment. We have demonstrated the effectiveness of RL meta-learning to create adaptive policies for aerospace applications in previous work \cite{gaudet2019seeker, gaudet2019adaptive,  gaudet2019guidance}.  In this work our goal is for the agent to hover at a position within 2m of its position at the start of the hovering maneuver, with constant attitude, and fuel expenditure minimized during hovering. Importantly, the optimized policy will be general in that it will allow hovering over \textit{any} asteroid with arbitrary shape, rotation, and density, provided the size is reasonably close to that of the synthetic asteroids used for optimization, and within the limits of thruster capability. To achieve this, the agent learns the policy in an environment that generates a new random asteroid for each episode.

The optimized policy serves as an integrated guidance, navigation, and control system for the purposes of a hovering maneuver, and interfaces with peripheral spacecraft systems as shown below in Fig. \ref{fig:policy_if}.

\begin{figure}[h!]
\begin{center}
\includegraphics[width=.95\linewidth]{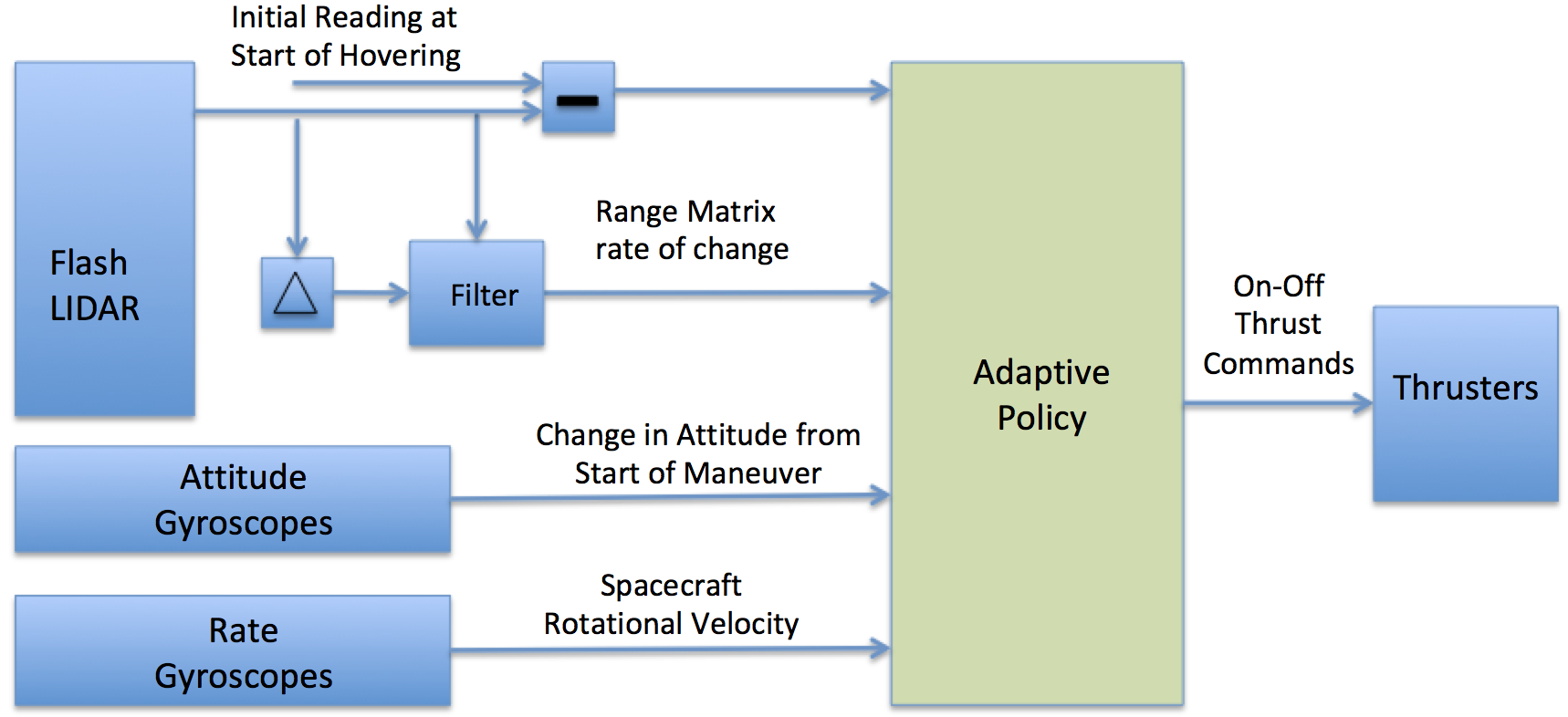}
\caption{Deployed Policy Interface with Peripheral Systems}
\label{fig:policy_if}
\end{center}
\end{figure}

\section{Problem Formulation}

\subsection{Spacecraft Configuration}

The spacecraft is modeled as a uniform density cube with height $h=2$, width $w=2$, and depth $d=2$, with inertia matrix given in Eq.~\eqref{eq:inertia_tensor}. The spacecraft has a wet mass ranging from 450 to 500 kg. The thruster configuration is shown in Table~\ref{tab:thrusters}, where $x$, $y$, and $z$ are the body frame axes.  Roll is about the $x$-axis, yaw is about the $z$-axis, and pitch is about the $y$-axis. Firing both thrusters on a face give translational thrust without rotation, while firing a single thruster on a given face induces a torque.  The navigation system provides updates to the guidance system every 6 s, and we integrate the equations of motion using fourth order Runge-Kutta integration with a time step of 2 s. Thrusters have a specific impulse of 210s.

\begin{equation}
    \label{eq:inertia_tensor}
    {\bf J}=\frac{m}{12}\begin{bmatrix} h^2 + d^2 & 0 & 0 \\ 0 & w^2+d^2 & 0 \\ 0 & 0 & w^2+h^2\end{bmatrix}
\end{equation}
$m$ is the spacecraft's mass, which is updated as shown in Eq.~\eqref{eq:EQOMc}.

\begin{table}[H]
	\fontsize{10}{10}\selectfont
    \caption{Body Frame Thruster Locations.}
   \label{tab:thrusters}
        \centering 
   \begin{tabular}{c | r | r | r | r | r} 
      \hline
      Thruster & x (m) & y (m) & z (m) & Thrust \\
      \hline
      1 & -1.0 & 0.0 & 0.4 & 1.0   \\
      2 & -1.0 & 0.0 & -0.4 & 1.0  \\
      3 & 1.0 & 0.0 & 0.4  & 1.0  \\
      4 & 1.0 & 0.0 & -0.4 & 1.0  \\
      5 & -0.4 & -1.0 & 0.0 & 1.0  \\
      6 & 0.4 & -1.0 & 0.0 & 1.0  \\
      7 & -0.4 & 1.0 & 0.0 & 1.0  \\
      8 & 0.4 & 1.0 & 0.0 & 1.0  \\
      9 & 0.0 & -0.4 & -1.0 & 1.0  \\
      10 & 0.0 & 0.4 & -1.0 & 1.0  \\
      11 & 0.0 & -0.4 & 1.0 & 1.0  \\
      12 & 0.0 & 0.4 & 1.0 & 1.0  \\
   \end{tabular}
\end{table}

\subsection{Asteroid and Sensor Models}

Since our goal is for the agent to hover above an asteroid with unknown shape, we need to insure that the agent does not learn the asteroid's shape during optimization.  To this end, we randomly generate a new asteroid for each episode.  Each asteroid starts as an icosahedron based on the unit sphere, after which we recursively (twice) expand each face into four equal triangles, with the new vertices projected onto the unit sphere. In the following, we will refer to this object as an "isosphere". Next, we randomly perturb each vertex of the unit isosphere by adding a value $\bf p \in \mathbb{R}^3$, where each element of $\bf p$ is uniformly drawn over the range $[-p,p]$, with $p$ uniformly drawn at the start of each episode from the range $[0.005, 0.05]$. Thus, different episodes will feature asteroids with different textures. We then randomly generate the asteroid's positive and negative $a$, $b$, and $c$ axes over the range 300 to 600 meters, and then scale the vertices appropriately. Since the positive and negative axes values are independently generated, this creates asymmetric $a$, $b$, and $c$ axes. A sample randomly generated asteroid with 1280 faces and 642 vertices is shown below in Fig. \ref{fig:rand_ast}.

\begin{figure}[h]
\begin{center}
\includegraphics[width=.9\linewidth]{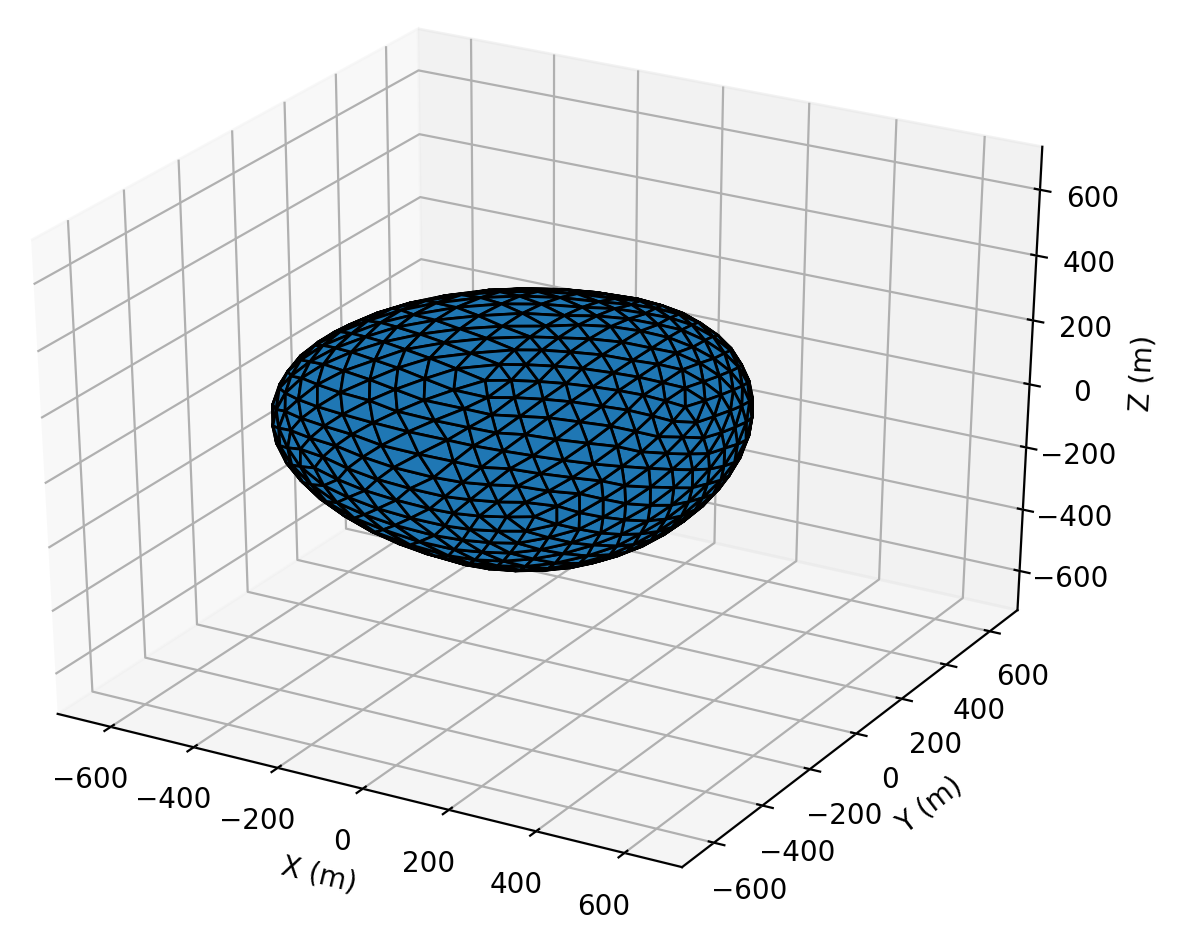}
\caption{Sample Random Asteroid}
\label{fig:rand_ast}
\end{center}
\end{figure}

For the modeling of environmental dynamics, we model the asteroid as an ellipsoid with uniform density. We assume that the asteroid is in general not rotating about a principal axis, and therefore to calculate the angular velocity vector we must specify the spin rate, the nutation angle (angle between the asteroid's z-axis and the axis of rotation), and moments of inertia \cite{scheeres2016orbital}.  The moments of inertia in turn depend on the asteroid's density and dimensions. The dimensions are specified by the ellipsoid axes $a = b\neq c$, where the axis constraints significantly simplifies the equations of motion. Since for the random asteroid the $a$ and $b$ axes are asymmetric and not in general equal, we average them for purposes of calculating the asteroid's rotational velocity components.  We use a gravity model that assumes a uniformly distributed sphere.  Although an ellipsoid model would have been more accurate, it would also be more computationally expensive, and with the asteroid sizes considered in this work, rotational forces dominate the dynamics. 

\begin{table}[h]
	\fontsize{10}{10}\selectfont
    \caption{Parameters for Randomly Generated Asteroids}
   \label{tab:ast_param}
        \centering 
   \newcolumntype{R}{>{\raggedleft\arraybackslash}p{2.5cm}}
   \begin{tabular}{l | R | R } 
        Parameter & min & max \\
       \hline
       a-axis (m) & 300 & 600 \\
       b-axis (m) & 300 & 600 \\
       c-axis (m) & 300 & 600 \\
      Mass $M$ (kg) & $1\times10^{10}$ & $150\time10^{10}$ \\
      Spin Rate $\omega_o$ (rad/s) & $5\times10^{-4}$  &  $1\times10^{-6}$ \\
      Nutation Angle (degrees) & 45 & 90 \\
      Acceleration due to SRP  $\text{m/s}^2$ &  $-100\times10^{-6}$  &  $100\times10^{-6}$
   \end{tabular}
\end{table}

The flash LIDAR is modeled as an 8 X 8 array of range sensors. Current commercial flash LIDAR units typically have a 100 X 100 sensor array, but the smaller sensor array allows much faster computation during optimization, and it seems intuitive that if hovering performance is satisfactory with an 8 X 8 array, it would likely be even better with a 100 X 100 array. In the unlikely event that this is not the case, an 8 X 8 array can be derived from a 100 X 100 array by downsampling. We use the Moller-Trumbore ray casting algorithm \cite{moller2005fast} to compute the intersection of each LIDAR beam with a triangle in the randomly generated asteroid. Our implementation only returns a range if a triangle is intersected on the correct side, and the intersection point is not occluded by another triangle. If a beam fails to intersect a triangle, a max range reading of 2000m is returned.

We also assume the spacecraft is equipped with a gyroscope that can measure the spacecraft's change in attitude (measured from the initiation of hovering), and a rate gyroscope to measure rotational velocity. At the beginning of an episode, the spacecraft's flash LIDAR is pointed in the same direction as the spacecraft's Z-axis. For the remainder of the episode, the LIDAR system is stabilized, i.e., we keep the LIDAR system's attitude constant during the maneuver as the spacecraft's attitude changes. Stabilization helps the hovering policy differentiate between changes in altimetry readings caused by rotation and changes caused by translation. In an actual implementation, this stabilization could be achieved by physically rotating the LIDAR platform to account for the spacecraft's change in attitude, similar to missile seeker stabilization \cite{siouris2004missile}.  We used a similar stabilization scheme for our work with seeker based guidance for asteroid close proximity operations \cite{gaudet2019seeker}.

\subsection{Equations of Motion}

The force $\mathbf{F}_{B}$ and torque $\mathbf{L}_{B}$ in the lander's body frame for a given commanded thrust depends on the placement of the thrusters in the lander structure. We can describe the placement of each thruster through a body-frame direction vector $\mathbf{d}$ and position vector $\mathbf{r}$, both in $\mathbb{R}^3$. The direction vector is a unit vector giving the direction of the body frame force that results when the thruster is fired.  The position vector gives the body frame location with respect to the  center of mass,  where the force resulting from the thruster firing is applied for purposes of computing torque, and in general the center of mass varies with time as fuel is consumed. For a lander with $k$ thrusters, the body frame force and torque associated with one or more  thrusters firing is then as shown in Equations \eqref{eq:Thruster_modela} and \eqref{eq:Thruster_modelb}, where $T_{cmd_{i}}\in[T_{min},T_{max}]$ is the commanded thrust for thruster $i$, $T_{min}$ and $T_{max}$ are a thruster's minimum and maximum thrust, $\mathbf{d}^{(i)}$ the direction vector for thruster $i$, and $\mathbf{r}^{(i)}$ the position of thruster $i$. The total body frame force and torque are calculated by summing the individual forces and torques.

\begin{subequations}
\begin{align}
	{\mathbf{F}_{B}}&={\sum_{i=1}^{k}\mathbf{d}^{(i)} T_{cmd}^{(i)}}\label{eq:Thruster_modela}\\
	{\mathbf{L}_{B}}&={\sum_{i=1}^{k}(\mathbf{r}^{(i)}-\mathbf{r}_\mathrm{com})\times\mathbf{F}_{B}^{(i)}}\label{eq:Thruster_modelb}
\end{align}
\end{subequations}

The dynamics model uses the lander's current attitude $\mathbf{q}$ to convert the body frame thrust vector to the inertial frame as shown in Equation \eqref{eq:BtoN} where $[\mathbf{BN}](\mathbf{q})$ is the direction cosine matrix mapping the inertial frame to body frame obtained from the current attitude parameter $\mathbf{q}$.

\begin{equation}
	\label{eq:BtoN}
	\mathbf{F}_{N}=\left[\left[\mathbf{BN}\right](\mathbf{q})\right]^{T}\mathbf{F}_{B}
\end{equation}

The rotational velocities $\bm{\omega}_{B/N}$ are then obtained by integrating the Euler rotational equations of motion, as shown in Equation \eqref{eq:EulerRot}, where $\mathbf{L}_{B}$ is the body frame torque as given in Equation \eqref{eq:Thruster_modelb}, $\mathbf{L}_{env}$ is the body frame torque from external disturbances, and $\mathbf{J}$ is the lander's inertia tensor. Note we have included a term that models a rotation induced by a changing inertia tensor.

\begin{equation}
	\label{eq:EulerRot}
	\mathbf{J}{\dot{\bm{\omega}}_{B}}=-\Tilde{\bm{\omega}}_{B}\mathbf{J}\bm{\omega}_{B}-\mathbf{\dot{J}}\bm{\omega}+\mathbf{L}_{B}+\mathbf{L}_{B_{env}}
\end{equation}

The lander's attitude is then updated by integrating the differential kinematic equations shown in Equation \eqref{eq:diffeqom}, where the lander's attitude is parameterized using the quaternion representation and $\bm{\omega}_{i}$ denotes the $i^{th}$ component of the rotational velocity vector $\bm{\omega}_{B}$. 

\begin{equation}
    \label{eq:diffeqom}
    \begin{bmatrix} \dot{q_{0}} \\ \dot{q_{1}} \\ \dot{q_{2}} \\ \dot{q_{3}}\end{bmatrix} = \frac{1}{2}\begin{bmatrix} q_{0} & -q_{1} & -q_{2} & -q_{3}\\ q_{1} & q_{0} & -q_{3} & q_{2}\\ q_{2} & q_{3} & q_{0} & -q_{1} \\ q_{3} & -q_{2} & q_{1} & q_{0} \end{bmatrix} \begin{bmatrix} 0 \\ \omega_{0} \\ \omega_{1} \\ \omega_{2} \end{bmatrix}
\end{equation}

The translational motion is modeled as shown in \ref{eq:EQOMa} through \ref{eq:EQOMc}.

\begin{subequations}
\begin{align}
	{\Dot{\mathbf r}} &= {{\mathbf v}}\label{eq:EQOMa}\\
	{\Dot{\bf v}} &= \frac{{{\bf F}_{N}}}{m} + {{\bf a}_\text{env}} - g(\mathbf{r},M) +2\mathbf{\dot{r}}\times\bm{\omega}_a + (\bm{\omega}_a\times\mathbf{r})\times\bm{\omega}_a\label{eq:EQOMb}\\
	\Dot{m} &= -\frac{\sum_{i}^{k}\lVert{{\bf F}_{B}}^{(i)}\rVert}{I_\text{sp}g_\text{ref}} \label{eq:EQOMc}
\end{align}
\end{subequations}
Here  ${{\bf F}_{N}}^{(i)}$ is the inertial frame force as given in Eq.~\eqref{eq:BtoN}, $k$ is the number of thrusters, $g_\text{ref}=9.8$ $\text{m}/\text{s}^{2}$,  $\mathbf{r}$ is the spacecraft's position in the asteroid centered reference frame,  $g(\mathbf{r},M)$ is a spherical gravity model and $M$ is the asteroid's mass, $I_\text{sp}=225$ s, and the spacecraft's mass is $m$.  ${\bf a}_\text{env}$ is a vector  representing solar radiation pressure. $\bm{\omega}_a$ is the asteroid's rotational velocity vector, which we compute as shown in Equations \eqref{eq:wa_a} through \eqref{eq:wa_f}, which uses the simplifying assumption that $J_x=J_y$ \cite{sawai2002control}. Here $\omega_o$ is the asteroid's spin rate and $\theta$ the nutation angle between the asteroid's spin axis and z-axis. We modified the equations from Reference (\citenum{sawai2002control}) to add the phase term $\phi$ to handle the case where the spacecraft starts the maneuver at an arbitrary point in the asteroid's rotational cycle. 

\begin{subequations}
\begin{align}
    \omega_{a_x} &= \omega_o\sin{\theta}\cos{(\omega_nt+\phi)}\label{eq:wa_a}\\
    \omega_{a_y} &= \omega_o\sin{\theta}\sin{(\omega_nt+\phi)}\label{eq:wa_b}\\
    \omega_{a_z} &= \omega_o\cos{\theta}\label{eq:wa_c}\\
    \omega_n &= \sigma\omega_o\cos{\theta}\label{eq:wa_d}\\
    J_{x,y}/J_z &= (b^2+c^2) / (a^2 + b^2)\label{eq:wa_e}\\
    \sigma &= \frac{(J_z-J_x)}{J_x}\label{eq:wa_f}
\end{align}
\end{subequations}

\section{Guidance Law Development}

\subsection{RL Overview}

In the RL framework, an agent learns through episodic interaction with an environment how to successfully complete a task by learning a policy  that maps observations to actions. The environment initializes an episode by randomly generating a ground truth state, mapping this state to an observation, and passing the observation to the agent. These observations could be a corrupted version of the ground truth state (to model sensor noise) or could be raw sensor outputs such as Doppler radar altimeter readings, a multi-channel pixel map from an electro-optical sensor, or in our case, a flash LIDAR range matrix.  The agent's policy uses this observation to generate an action that is sent to the environment; the environment then uses the action and the current ground truth state to generate the next state and a scalar reward signal.  The reward and the observation corresponding to the next state are then passed back to the agent. The process repeats until the environment terminates the episode, with the termination signaled to the agent via a done signal. Possible termination conditions include the agent completing the task, satisfying some condition on the ground truth state (such as altitude falling below zero), or violating a constraint.  
 
A Markov Decision Process (MDP) is an abstraction of the environment, which in a continuous state and action space, can be represented by a state space $\mathcal{S}$, an action space $\mathcal{A}$, a state transition distribution $\mathcal{P}(\mathbf{x}_{t+1}|\mathbf{x}_t,\mathbf{u}_t)$, and a reward function $r=\mathcal{R}(\mathbf{x},\mathbf{u}))$, where $\mathbf{x} \in \mathcal{S}$ and $\mathbf{u} \in \mathcal{A}$, and $r$ is a scalar reward signal. We can also define a partially observable MDP (POMDP), where the state $\mathbf{x}$ becomes a hidden state, generating an observation $\mathbf{o}$ using an observation function $\mathcal{O}(\mathbf{x})$ that maps states to observations. The POMDP formulation is useful when the observation consists of raw sensor outputs, as is the case in this work.  In the following, we will refer to both fully observable and partially observable environments as POMDPs, as an MDP can be considered a POMDP with an identity function mapping states to observations.

The agent operates within an  environment defined by the POMDP, generating some action $\mathbf{u}_t$ based off of the observation $\mathbf{o}_t$, and receiving reward $r_{t+1}$ and next observation $\mathbf{o}_{t+1}$. Optimization involves maximizing the sum of (potentially discounted) rewards over the trajectories induced by the interaction between the agent and environment. Constraints such as minimum and maximum thrust, glide slope, attitude compatible with sensor field of view,  maximum rotational velocity, and terrain feature avoidance (such as targeting the bottom of a deep crater) can be included in the reward function, and will be accounted for when the policy is optimized. Note that there is no guarantee on the optimality of trajectories induced by the policy, although in practice it is possible to get close to optimal performance by tuning the reward function \cite{gaudet2018deep}.

Reinforcement meta-learning differs from generic reinforcement learning in that the agent learns to quickly adapt to novel POMDPs by learning over a wide range of POMDPs. These POMDPs can include different environmental dynamics, actuator failure scenarios, mass and inertia tensor variation, and varying amounts of sensor distortion. Learning within the RL meta-learning framework results in an agent that can quickly adapt to novel POMDPs, often with just a few steps of interaction with the environment. There are multiple approaches to implementing meta-RL.  In \cite{finn2017model}, the authors design the objective function to explicitly make the model parameters transfer well to new tasks, whereas in \cite{mishra2017simple} the authors demonstrate state of the art performance using temporal convolutions with soft attention. And  in \cite{frans2017meta}, the authors use a hierarchy of policies to achieve meta-RL. In this proposal, we use a different approach \cite{wang2016learning} using a recurrent policy and value function. Note that it is possible to train over a wide range of POMDPs using a non-meta RL algorithm \cite{gaudet2018deep,rajeswaran2016epopt}. Although such an approach typically results in a robust policy, the policy cannot adapt in real time to novel environments. 

In this work, we  implement metal-RL using proximal policy optimization (PPO) \cite{schulman2017proximal} with both the policy and value function implementing recurrent layers in their networks.  To understand how recurrent layers result in an adaptive agent, consider that given some ground truth agent state $\mathbf{x}_{t}$ and action vector $\mathbf{u}_{t}$ output by the agent's policy, the next state $\mathbf{x}_{t+1}$ and observation $\mathbf{o}_{t+1}$ depends not only on $\mathbf{x}_{t}$ and $\mathbf{u}_{t}$, but also on the ground truth agent mass, inertia tensor, and external forces acting on the agent, as well as the asteroid's shape. Specifically, during optimization, the hidden state of a network's recurrent network evolves differently depending on the observed sequence of observations from the environment and actions output by the policy, with the state evolution capturing unobserved and potentially time-varying information, such as external forces, that are useful in minimizing the cost function.  In contrast, a non-recurrent policy, which does not maintain a persistent hidden state vector, can only optimize using a set of current observations, actions, and advantages, and will tend to under-perform a recurrent policy on tasks with randomized dynamics \cite{gaudet2019adaptive}.  After training, although the recurrent policy's network weights are frozen, the hidden state will continue to evolve in response to a sequence of observations and actions, thus making the policy adaptive.  In contrast, a policy without a recurrent network layer has behavior that is fixed by the network parameters at test time.

The PPO algorithm used in this work  is a  policy gradient algorithm which has demonstrated state-of-the-art performance for many RL benchmark problems. PPO approximates the Trust Region Policy Optimization (TRPO) process \cite{schulman2015trust} by accounting for the policy adjustment constraint with a clipped objective function. The objective function used with PPO can be expressed in terms of the probability ratio $p_{k}({\bm \theta})$ given by Eq.~\eqref{eq:clipr}, where $\pi_\theta$ is the policy parameterized by parameter vector $\theta$.
\begin{equation}
\label{eq:clipr} 
p_{k}({\bm \theta})=\frac{\pi_{{\bm \theta}}({\bf u}_{k}|{\bf o}_{k})}{\pi_{{\bm \theta}_\text{old}}({\bf u}_{k}|{\bf o}_{k})}
\end{equation}
The PPO objective function is shown in Equations ~\eqref{eq:ppoloss_a} through ~\eqref{eq:ppoloss_c}.  The general idea is to create two surrogate objectives, the first being the probability ratio $p_{k}({\bm \theta})$ multiplied by the advantages $A^{\pi}_{\bf w}({\bf o}_{k},{\bf u}_{k})$ (see Eq. \eqref{eq:ppo_adv}), and the second a clipped (using clipping parameter $\epsilon$) version of $p_{k}({\bm \theta})$ multiplied by $A^{\pi}_{\bf w}({\bf o}_{k},{\bf u}_{k})$.  The objective to be maximized $J({\bm \theta})$ is then the expectation under the trajectories induced by the policy of the lesser of these two surrogate objectives.
\begin{subequations}
\begin{align}
	\text{obj1} &= p_{k}({\bm \theta})A^{\pi}_{\bf w}({\bf o}_{k},{\bf u}_{k})\label{eq:ppoloss_a}\\
	\text{obj2} &= \mathrm{clip}(p_{k}({\bm \theta})A^{\pi}_{\bf w}({\bf o}_{k},{\bf u}_{k}) , 1-\epsilon, 1+\epsilon)\label{eq:ppoloss_b}\\
	J({\bm \theta})&=\mathbb{E}_{p({\bm \tau})}[\mathrm{min}(\text{obj1},\text{obj2})]\label{eq:ppoloss_c}
\end{align}
\end{subequations}

This clipped objective function has been shown to maintain a bounded KL divergence with respect to the policy distributions between updates, which aids convergence by insuring that the policy does not change drastically between updates. Our implementation of PPO uses an approximation to the advantage function that is the difference between the empirical return and a state value function baseline, as shown in Equation \ref{eq:ppo_adv}:
\begin{equation}
\label{eq:ppo_adv}
    A^{\pi}_{\bf w}(\mathbf{o}_{k},\mathbf{u}_{k})=\left[\sum_{\ell=k}^{T}\gamma^{\ell-k}r(\bf o_{\ell},\bf u_{\ell})\right]-V_{\bf w}^{\pi}(\mathbf{o}_{k})
\end{equation}
Here the value function $V_{\bf w}^{\pi}$ parameterized by vector $\mathbf w$ is learned using the cost function given by Eq.~\eqref{eq:vf_ppo}, where $\gamma$ is a discount rate applied to rewards generated by reward function $\mathcal{R}(\mathbf{o},\mathbf{u})$.  The discounting of rewards improves optimization performance by improving temporal credit assignment.
\begin{equation}
\label{eq:vf_ppo}
L(\mathbf{w})=\sum_{i=1}^{M}\left(V_{\mathbf{w}}^{\pi}({\bf o}_k^i)-\left[\sum_{\ell=k}^{T}\gamma^{\ell-k}\mathcal{R}({\bf u}_{\ell}^i,{\bf o}_{\ell}^i)\right]\right)^2
\end{equation}
In practice, policy gradient algorithms update the policy using a batch of trajectories (roll-outs) collected by interaction with the environment. Each trajectory is associated with a single episode, with a sample from a trajectory collected at step $k$ consisting of observation ${\bf o}_{k}$, action ${\bf u}_{k}$, and reward $r_k=\mathcal{R}({\bf o}_k,{\bf u}_k)$. Finally, gradient ascent is performed on ${\bm \theta}$ and gradient descent on ${\bf w}$ and update equations are given by
\begin{align}\label{loss}
{\bf w}^+&={\bf w}^--\beta_{{\bf w}}\nabla_{{\bf w}} \left. L({\bf w})\right|_{{\bf w}={\bf w}^-}\\
{\bm \theta}^+&={\bm \theta}^-+\beta_{{\bm \theta}} \left. \nabla_{\bm \theta}J\left({\bm \theta}\right)\right|_{{\bm \theta}={\bm \theta}^-}
\end{align}
where $\beta_{{\bf w}}$ and $\beta_{{\bm \theta}}$ are the learning rates for the value function, $V_{\bf w}^{\pi}\left({\bf o}_k\right)$, and policy, $\pi_{\bm \theta}\left({\bf u}_k|{\bf o}_k\right)$, respectively.

In our implementation, we dynamically adjust the clipping parameter $\epsilon$ to target a KL divergence between policy updates of 0.001. The policy and value function are learned concurrently, as the estimated value of a state is policy dependent. The policy uses a multi-categorical policy distribution, where a separate observation conditional categorical distribution is maintained for each element of the action vector. Note that exploration in this case is conditioned on the observation, with the two logits associated with each element of the action vector determining how peaked the softmax distribution becomes for each action. Because the log probabilities are calculated using the logits, the degree of exploration automatically adapts during learning such that the objective function is maximized. Finally, note that a full categorical distribution would be impractical, as the number of labels would be $2^{12}$, as opposed to $2 \times 12$ for the multi-categorical distribution.

\subsection{Guidance Law Optimization}

\begin{figure}[h]
\begin{center}
\includegraphics[width=.9\linewidth]{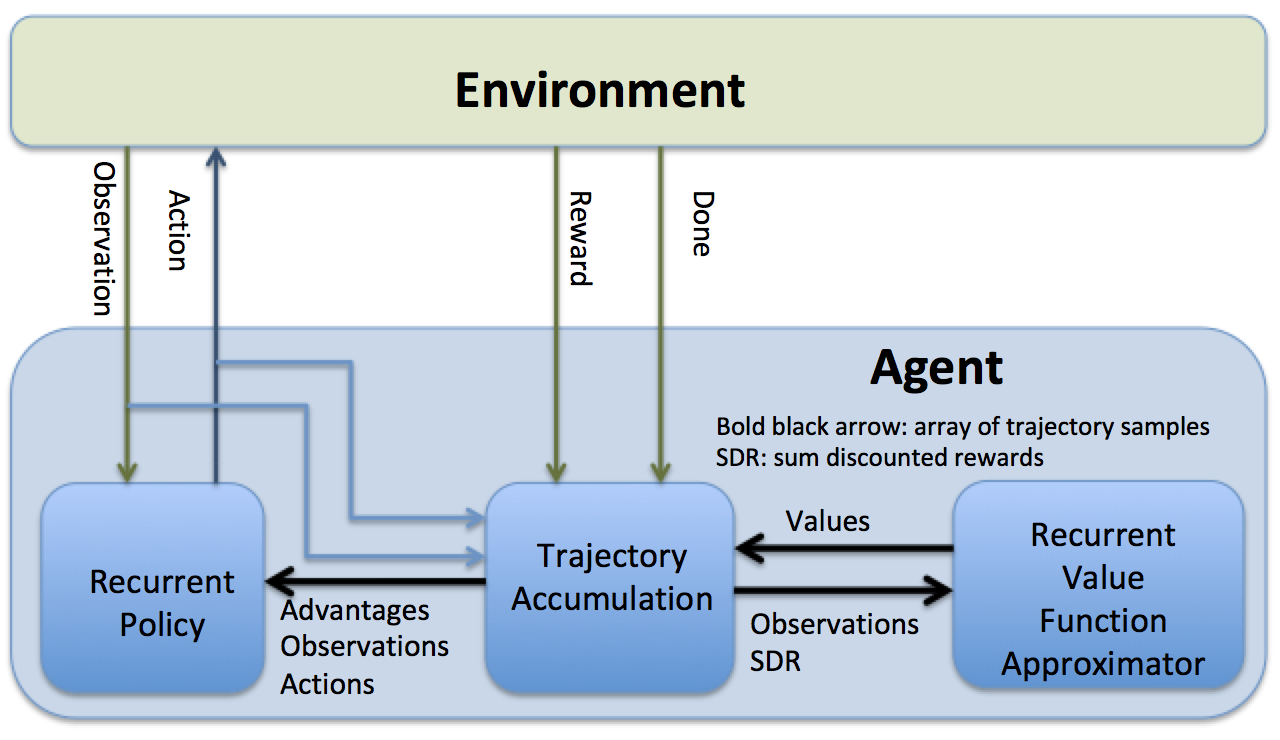}
\caption{Agent-Environment Interface}
\label{fig:a_e_i}
\end{center}
\end{figure}

A simplified view of the agent and environment are shown in Figure \ref{fig:a_e_i}. The environment instantiates the system dynamics model, asteroid shape model, reward function, spacecraft model, and thruster model. Note that when using a policy gradient method such as PPO it suffices to deploy the policy, and it is not necessary to deploy the value function.  We can take advantage of this by giving the value function access to the ground truth state during optimization, whereas the policy only has access to the observations, in this case the flash LIDAR measurements. Specifically, the value function has access to the observation given in Eq. \eqref{eq:vf_obs}, where $\bf r_{\text{err}}$ and $\bf{dq}$ are the changes in the agent's position and attitude since the initiation of the hovering maneuver, $\bf v$ is the agent's velocity, and $\boldsymbol \omega$ the spacecraft's rotational velocity. 

\begin{equation}
    \label{eq:vf_obs}
    \mathrm{obs_{VF}} = \begin{bmatrix} \bf{r}_{\text{err}} & \bf v & \boldsymbol{dq} & \boldsymbol{\omega} &\end{bmatrix} 
\end{equation}

On the other hand, the policy only has access to the difference between the matrix of flash LIDAR readings at the current timestep and the readings at the start of the hovering maneuver $\bf R_{err}$, the change in LIDAR readings between consecutive measurements $\bf{dR}$, along with the estimated rotational velocity $\boldsymbol{\omega}$ and change in attitude since the start of the hovering maneuver $\boldsymbol{dq}$. Using $\bf R_{err}$ as opposed to the actual range matrix $\text R$ allows the agent to generalize better to different altitude ranges.  Note that in an actual implementation, $\bf{dR}$ would be smoothed with a Kalman filter. The observation given to the policy is then as shown in Eq.~\eqref{eq:obs}.

\begin{equation}
    \label{eq:obs}
    \mathrm{obs_{\pi}} = \begin{bmatrix} \bf{R_{err}} & \bf{dR} & \boldsymbol{dq} & \boldsymbol{\omega} &\end{bmatrix} 
\end{equation}

The action space is in $\mathbb{Z}^{k}$, where $k$ is the number of thrusters.  Each element of the agent action $\mathbf{u} \in {0,1}$ is used to index Table \ref{tab:thrusters}, where if the  action is 1, it is used to compute the body frame force and torque contributed by that thruster.   

The value function is implemented using a four layer neural network with tanh activations on each hidden layer. Layer 2 for the  value function network is a recurrent layer implemented as a gated recurrent unit \cite{chung2015gated}. The network architecture is  as shown in Table \ref{tab:VFNN}, where $n_{\mathrm{hi}}$ is the number of units in layer $i$ and $\mathrm{obs\_dim}$ is the observation dimension 

\begin{table}[h]
	\fontsize{10}{10}\selectfont
    \caption{Value Function network architecture}
   \label{tab:VFNN}
        \centering 
   \newcolumntype{R}{>{\raggedleft\arraybackslash}p{1.8cm}}
   \begin{tabular}{l | R | c  } 
       \hline
       Layer & \# units & activation  \\
       \hline
      hidden 1      & $10 * \mathrm{obs\_dim}$ & tanh \\
      hidden 2      &  $\sqrt{n_{\mathrm{h1}} * n_{\mathrm{h3}}}$ & tanh\\
      hidden 3      &  5 & tanh \\
      output        &  1 & linear \\
      \hline
   \end{tabular}
\end{table}

\begin{figure}[h]
\begin{center}
\includegraphics[width=.5\linewidth]{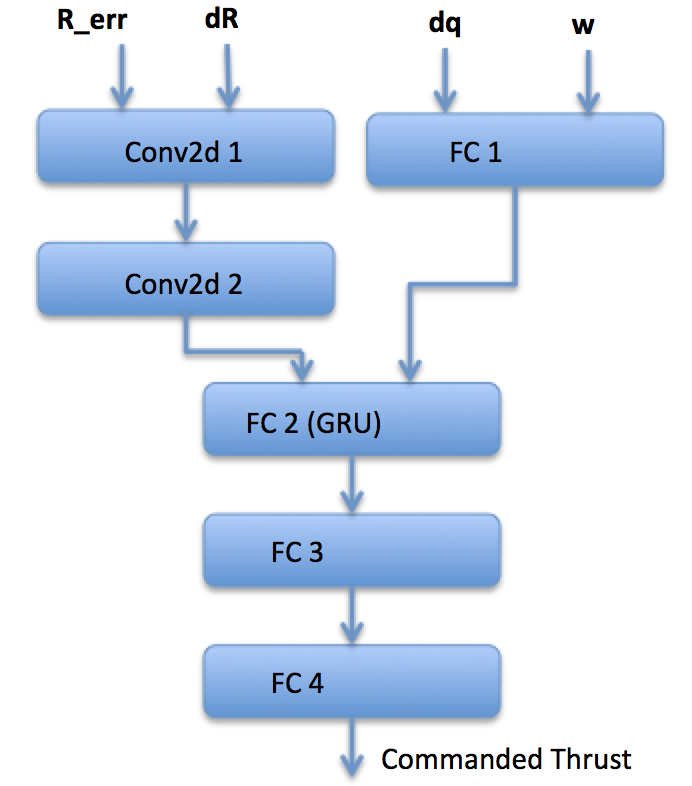}
\caption{Policy Network}
\label{fig:policy_net}
\end{center}
\end{figure}

The policy has a convolutional \cite{lecun1995convolutional} front end with an architecture inspired by [\citenum{springenberg2014striving}], where the pooling layer is replaced by a 2-D convolutional layer with stride 2. We have found that this improves performance for RL applications. We use rectified linear activations units for each convolutional layer.  The first convolutional layer has 2 channels (one for the range readings, the other for the difference in range readings), 8 filters, a filter size of 3, and stride of 1. The second convolutional layer has 8 channels and 8 filters, a filter size of 4, and a stride of 2. The final layers are fully connected, as shown in \ref{tab:PNN}.  The entire policy network is diagrammed in Fig. \ref{fig:policy_net}. The policy and value functions are periodically updated during optimization after accumulating trajectory rollouts of 30 simulated episodes.
\begin{table}[h]
	\fontsize{10}{10}\selectfont
    \caption{Fully Connected Policy network layers}
   \label{tab:PNN}
        \centering 
   \newcolumntype{R}{>{\raggedleft\arraybackslash}p{1.8cm}}
   \begin{tabular}{l | R | c  } 
       \hline
       Layer & \# units & activation  \\
       \hline
      FC 1      & 70 & tanh \\
      FC 2      & 154 & tanh \\
      FC 3      & 120 & tanh  \\
      FC 4        & 12 & linear  \\
      \hline
   \end{tabular}
\end{table}

During optimization, the agent is given negative rewards proportional to the cumulative change in position from the start of the hovering maneuver. Large negative rewards are given for exceeding a maximum rotational velocity of 0.10 rad/s or if the attitude is such that all of the flash LIDAR elements miss the asteroid, which is detected by all elements returning a max range reading of 2000m. Constraint violation also results in the termination of the current episode. Small negative rewards are given proportional to the control effort at each timestep. 

Finally, we provide a terminal reward bonus when the spacecraft executes a good landing (see below). The reward function is then given by Equation \eqref{eq:reward_func}, where the various terms are described in the following:

\begin{enumerate}
    \item $\alpha$ weights a term penalizing the current deviation from desired hovering position.
    \item $\beta$ weights a term penalizing deviation from desired hovering attitude.
    \item $\gamma$ weights a term penalizing control effort.
    \item $\eta$ is a constant positive term that encourages the agent to keep making progress along the trajectory.
    \item $\zeta$ is a bonus given for satisfying a terminal constraint at the end of the hovering maneuver, where the spacecraft's terminal position and velocity are all within specified limits.  The limits are $\|\mathbf{r}\|=2$ $\text{m}$, $\|\mathbf{v}\|=0.1$, $\text{m/s}$, and all components of angular velocity less than 0.025 rad/sec
    \item $\kappa$ is a penalty for exceeding any constraint. We impose a rotational velocity constraint of 0.10 rad/sec for all three rotational axes. We also constrain the spacecraft's attitude such that at least one LIDAR beam hits the asteroid.  If all beams miss the asteroid, we assume the attitude constraint is violated.
\end{enumerate}

\begin{equation}
 \begin{split}
 \label{eq:reward_func}
r &= \alpha r_{\text err}+ \beta q_{\text error} + \gamma\|{\bf T}\|+ \eta + \\&\zeta(\text{terminal  constraints satisfied})+\kappa(\text{constraint violation})
\end{split}
\end{equation}

Initial hyperparameter settings are shown in Table \ref{tab:HPS}.

\begin{table}[h]
	\fontsize{10}{10}\selectfont
    \caption{Hyperparameter Settings}
   \label{tab:HPS}
        \centering 
   \begin{tabular}{ c | c | c | c | c | c  } 
      \hline
      $\alpha$  & $\beta$   &  $\gamma$ & $\eta$ & $\zeta$ & $\kappa$\\
      \hline
       -0.02 & -0.01 & -0.05 & 0.01 & 10 & -50\\
      
   \end{tabular}
\end{table}

\section{Experiments}

Code to reproduce these experiments will be made available on our Github page\footnote{https://github.com/Aerospace-AI/Aerospace-AI.github.io}. The spacecraft initial condition limits for these experiments were selected assuming that the spacecraft would start with its sensor pointed in the general direction of the asteroid with minimal residual translational and rotational velocities.  The initial conditions used for our experiments are given in Table \ref{tab:initial_conditions}.  Note that the initial range is with respect to the asteroid's surface given the line of sight to the asteroid center from the spacecraft's initial position, i.e., a range of 100m implies an altitude of 100m with respect to the asteroid's surface, regardless of the asteroid dimensions. Position $\theta$ and $\phi$ along with the initial range (plus the asteroid radius where it is collinear with the initial line of sight) specify the spacecraft's position in spherical coordinates in the asteroid centered reference frame.   The spacecraft has a small uniformly distributed initial velocity. 

The spacecraft's ideal initial attitude is such that the -Z body-frame axis is aligned with the line of sight to target. This ideal initial attitude is perturbed at the start of each episode such that the angle between the -Z body frame axis and line of sight to target varies uniformly as shown in Table \ref{tab:initial_conditions}.  Lower  hovering altitudes and larger asteroids both give rise to a scenario where  most of the flash LIDAR beams give valid returns, and under these conditions the guidance algorithm can tolerate larger initial attitude errors than that shown in Table \ref{tab:initial_conditions}. Similarly, the guidance system can tolerate higher initial altitudes for smaller initial attitude errors and larger asteroids. At the start of each episode, a slight actuator failure is deemed to occur with probability 0.5.  This actuator failure results in the thrust for a randomly chosen thruster to be reduced by a factor of 0.9. 

\begin{table}[h]
	\fontsize{10}{10}\selectfont
    \caption{Initial Conditions}
   \label{tab:initial_conditions}
        \centering 
   \newcolumntype{R}{>{\raggedleft\arraybackslash}p{2cm}}
   \begin{tabular}{l | R | R } 
        Parameter & min & max \\
       \hline
       Range (m) & 100.0 & 600.0 \\
      Position $\theta$ (degrees) & 0.0 & 90.0  \\
      Position $\phi$ (degrees) & $-\pi$ & $\pi$ \\
      x comp of Velocity (cm/s) & -10.0 & 10.0 \\
      y comp of Velocity (cm/s) & -10.0 & 10.0 \\
      z comp of Velocity (cm/s) & -10.0 & 10.0 \\
      Attitude Error (degrees) & 0.0 & 11.0 \\
      x comp of Rotational Velocity (mrad/s) & -20.0 & 20.0 \\
      y comp of Rotational Velocity (mrad/s) & -20.0 & 20.0 \\
      z comp of Rotational Velocity (mrad/s) & -20.0 & 20.0 \\
   \end{tabular}
\end{table}

\subsection{Optimization Results}

\begin{figure}[h]
\begin{center}
\includegraphics[width=.9\linewidth]{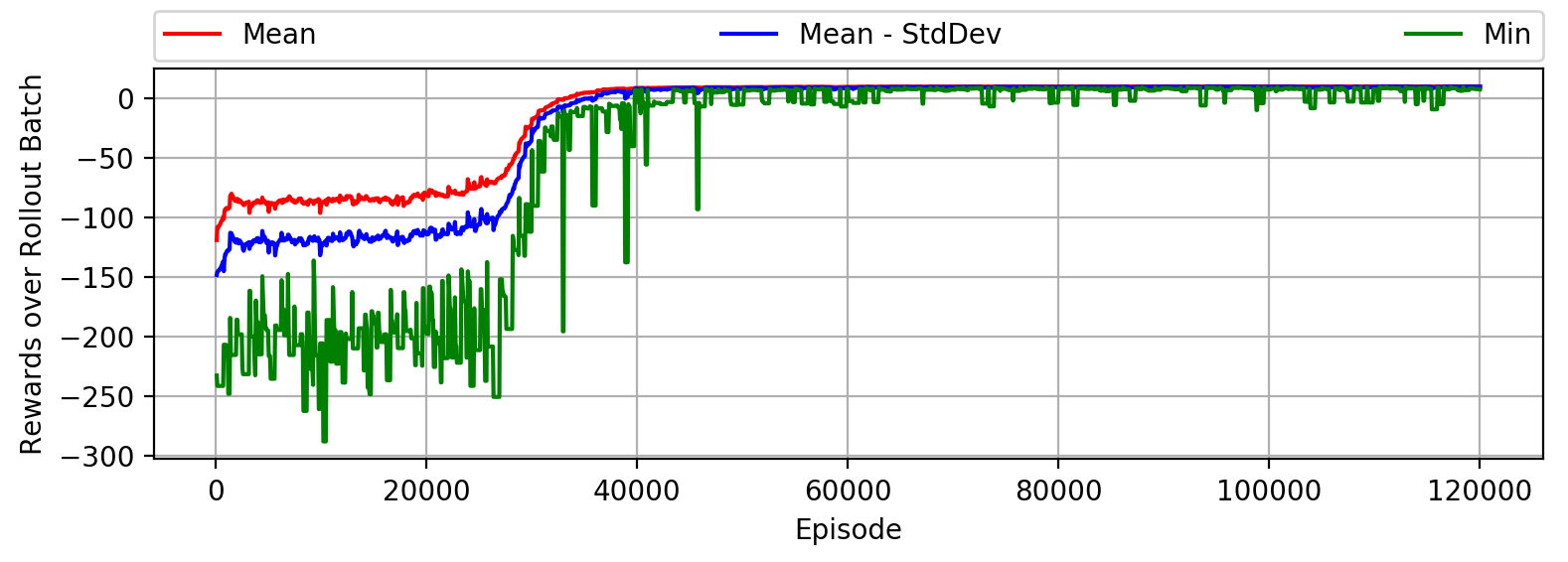}
\caption{Optimization Rewards Learning Curves}
\label{fig:rewards}
\end{center}
\end{figure}

\begin{figure}[h]
\begin{center}
\includegraphics[width=.9\linewidth]{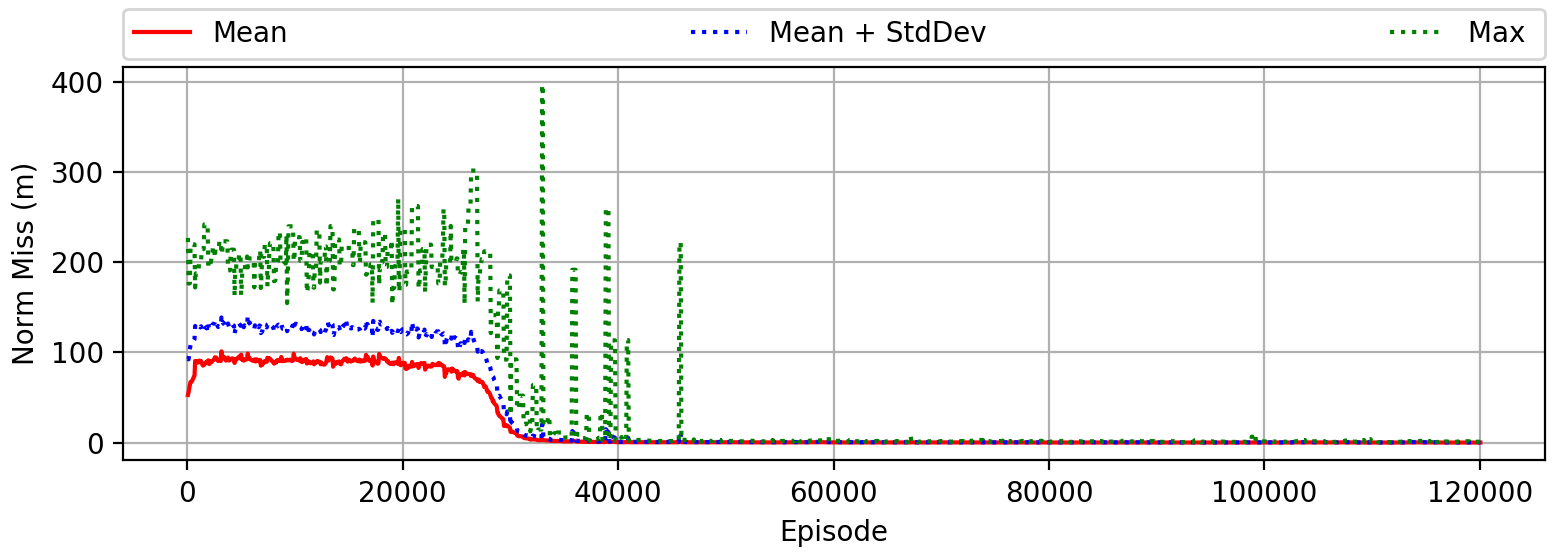}
\caption{Optimization Terminal Position Error Learning Curves}
\label{fig:rf}
\end{center}
\end{figure}

We optimize using 30 episode rollouts and the initial conditions given in Table \ref{tab:initial_conditions}. To reduce computational requirements, we use an asteroid with only 320 facets for optimization. Each episode attempts to hover for 600s, but early termination is possible in the event of a constraint violation. For each episode, we randomly generate a new asteroid using parameters as given in Table \ref{tab:ast_param}.  Fig.~\ref{fig:rewards}  plots reward statistics  and Fig.~\ref{fig:rf} plots the terminal position error statistics, with statistics for both plots computed over rollout batch of 30 episodes.  We see that initially the position error is high as the policy is focusing on satisfying the constraints that at least one element of the flash LIDAR sensor returns a valid reading and the maximum rotational velocity is not exceeded. Once the policy learns to satisfy the constraints, it focuses on minimizing the position error.  

\subsection{Policy Testing: Synthetic Asteroids}

\begin{figure}[h!]
\begin{center}
\includegraphics[width=1.0\linewidth]{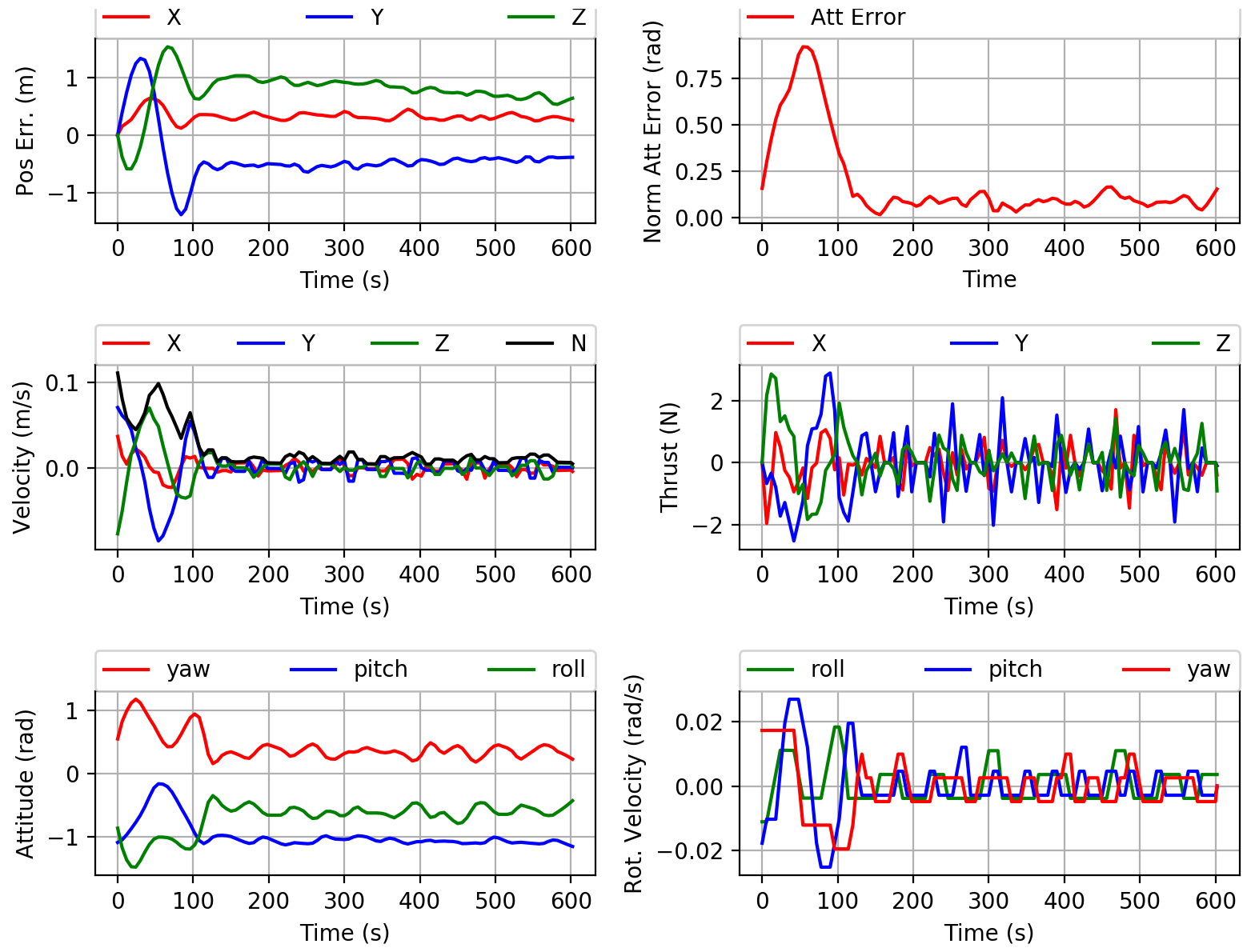}
\caption{Sample trajectory}
\label{fig:traj}
\end{center}
\end{figure}

We begin by testing the optimized policy on randomly generated synthetic asteroids, using the same initial conditions and asteroid parameters as in optimization. Note that unique scenarios are encountered in testing due to the random selection of asteroid and initial condition parameters.   Test results are shown in Table \ref{tab:results}, which are computed from 5000 simulated episodes. Note that the rotational velocity row in Table \ref{tab:results} gives the rotational velocity vector element with the worst performance (highest absolute value). The "Good Hover 1" row gives the percentage of episodes where the terminal position error was less than 2m, terminal speed less than 10cm/s, and all elements of the terminal rotational velocity less than 0.015 rad/s. The "Good Hover 2" row has the same terminal speed and rotational velocity constraints, but only requires the terminal position error to be less than 5m. We achieved our terminal performance goals in all 5000 episodes. As a back of envelope calculation for the fuel required to hover, the mean force acting on the spacecraft (rotational and gravitational) was 0.5N. Plugging in the spacecraft's specific impulse and 300 simulated steps over the 600s hovering duration, we find that 0.08kg of fuel would be required to cancel the environmental forces. Our actual average fuel consumption of 0.41kg is considerably larger. Part of the excess fuel consumption can be attributed to using pulsed thrusters, which do not allow exact cancellation of environmental forces.  It is also possible we could have increased fuel efficiency by using a higher (absolute) value for the $\beta$ coefficient in Table \ref{tab:HPS}.   A sample trajectory is shown in Figure \ref{fig:traj}, where the position error subplot plots the deviation from the spacecraft's initial position. Note that when the environmental dynamics are such that maximum thrust is not required, the policy fires only a single thruster on a given side of the spacecraft, resulting in a 1N thrust. This saves fuel at the expense of inducing rotation, which is compensated for by firing the opposing thruster on the opposite side at some future time.

To illustrate the ability of the policy to generalize to novel scenarios, we re-ran testing using the cases tabulated in Table \ref{tab:gen_cases}.  Except as noted in this table, the initial conditions and asteroid characteristics were identical to that used for optimization. In each case, performance was similar but slightly worse to that shown in Table \ref{tab:results}, with the exception of increased fuel consumption for the extended hovering duration test, and rare failures to achieve the required hovering performance. Abbreviated results are given in Table \ref{tab:ext_results}, where we see that the policy has trouble generalizing to lower altitudes. This may be due to the reduced perceived curvature at lower altitudes, and an obvious remedy that could be explored in future work would be to optimize over positions that cover these lower altitudes.  We ran additional experiments with a minimum altitude of 50m that resulted in performance closer to that of Table \ref{tab:results}. The policy generalized fairly well to  a more finely grained texture (more facets).  

\begin{table}[h!]
	\fontsize{10}{10}\selectfont
    \caption{Performance}
   \label{tab:results}
        \centering 
   \newcolumntype{R}{>{\raggedleft\arraybackslash}p{1.5cm}}
   \begin{tabular}{ l | R | R | R   } 
      \hline
      Parameter & Mean & Std & Max \\
      \hline
      Terminal Position (m) & 0.21 & 0.15 & 1.87 \\
      Terminal Velocity (cm/s) & 0.9 & 0.4 & 3.3\\
      Rotational Velocity (mrad/s) & 0.00 & 0.46 & 1.47 \\
      Good Hover 1 (\%) & 100.0 & N/A & N/A \\
      Good Hover 2 (\%) & 100.0 & N/A & N/A \\
      Fuel (kg) & 0.41 & 0.09 & 0.63 \\
   \end{tabular}
\end{table}

\begin{table}[h!]
	\fontsize{10}{10}\selectfont
    \caption{Generalization Cases}
   \label{tab:gen_cases}
        \centering 
   \renewcommand{\arraystretch}{1.5}
   \begin{tabular}{p{4.0cm} | p{6.0cm}  } 
 
      \hline
      Case & Description \\
      \hline
      Extended Altitude Range & Initial altitude range increased to (10m, 700m)  \\
      Asteroid Facets & Number of facets on each randomly generated asteroid shape model increased from 320 to 1280\\
      Hovering Duration & We increased hovering duration to 1200s \\
      Actuator Failure & With probability 0.5, random failing thruster has thrust reduced by factor of 0.5 \\
      Sensor Noise & Random range bias for each episode uniformly distributed between -5m and 5m, Gaussian zero mean range noise with 2m standard deviation at each sample \\
      Environmental Dynamics & Maximum asteroid spin rate increased to 1e-3 rad/s \\
      Center of Mass (COM) Variation & At the start of each episode, the spacecraft's center of mass is randomly set to an initial value between -10cm and 10cm on each axis.
      
   \end{tabular}
\end{table}

\begin{table}[h!]
	\fontsize{10}{10}\selectfont
    \caption{Generalization Performance}
   \label{tab:ext_results}
        \centering 
   \newcolumntype{R}{>{\raggedleft\arraybackslash}p{2.5cm}}
   \begin{tabular}{ l | R | R | R   } 
      \hline
      Case & Good Hover 1 (\%) & Good Hover 2 (\%) & Max Pos Err (m)\\
      \hline
      Extended IC & 98.22 & 99.76 & 25.2\\
      Facets & 100.00 & 99.94 & 4.46 \\
      Duration & 99.96  & 100.00 & 3.73 \\
      Actuator Fail  & 100.00 & 100.00 & 1.66\\
      Sensor Noise  & 100.00 & 98.86 &  3.67\\
      Env. Dynamics & 100.00 & 100.00 & 1.78 \\
      COM Variation & 100.00 & 100.00 & 1.37 \\
      
   \end{tabular}
\end{table}


\subsection{Policy Testing: rq36 and Itokawa}

Since a guidance  law for hovering over synthetic asteroids would be of limited value, we also test the optimized policy using a shape model of asteroids rq36 and Itokawa. These shape models are shown below in Fig. \ref{fig:rq36_and_itokawa3x}, along with flash LIDAR beams (green for hit, red for miss) from a randomly generated spacecraft initial state. Due to the slightly smaller size of asteroid rq36 as compared to that of the  smallest randomly generated asteroids, rare complete misses for the flash LIDAR returns occurred when the altitude was above 500m. We define a complete miss as none of the flash LIDAR beams intersecting the asteroid. Similarly, the minimum dimension of the peanut shaped asteroid Itokawa also resulted in occasional complete misses at altitudes greater than 250m when the spacecraft was located close to collinear with the asteroid's x-axis. Since the curvature of Itokawa is quite different from that of the synthetic asteroids, it should be a particularly challenging test case, and we therefore look at two cases.  First, we test using the standard Itokawa shape model, but restrict the altitude to below 250m. Second, we scaled up the dimensions of the Itokawa shape model by a factor of 3, which we refer to in the following as "Itokawa3X". This scaling resulted in a minimum axis size slightly larger than the smallest experienced during optimization, which allowed  testing hovering at higher altitudes.  Other than the modified initial altitude shown in Table \ref{tab:real_IC}, the initial conditions are identical to that for the random synthetic asteroid testing. A performance summary is given in Table \ref{tab:real_results}, where we see that, similar to the case of the synthetic asteroids, the policy has trouble generalizing to hovering at low altitudes (below 100m), particularly for the case of Itokawa3X. We ran additional experiments and found that hovering at a minimum altitude of 20m gives performance close to that observed at 100m. Also note that the performance of Itokawa3X is a bit worse than the other cases, perhaps due to the policy failing to generalize to the smaller curvature along the major axis.

\begin{figure}[h!]
\begin{center}
\includegraphics[width=1.0\linewidth]{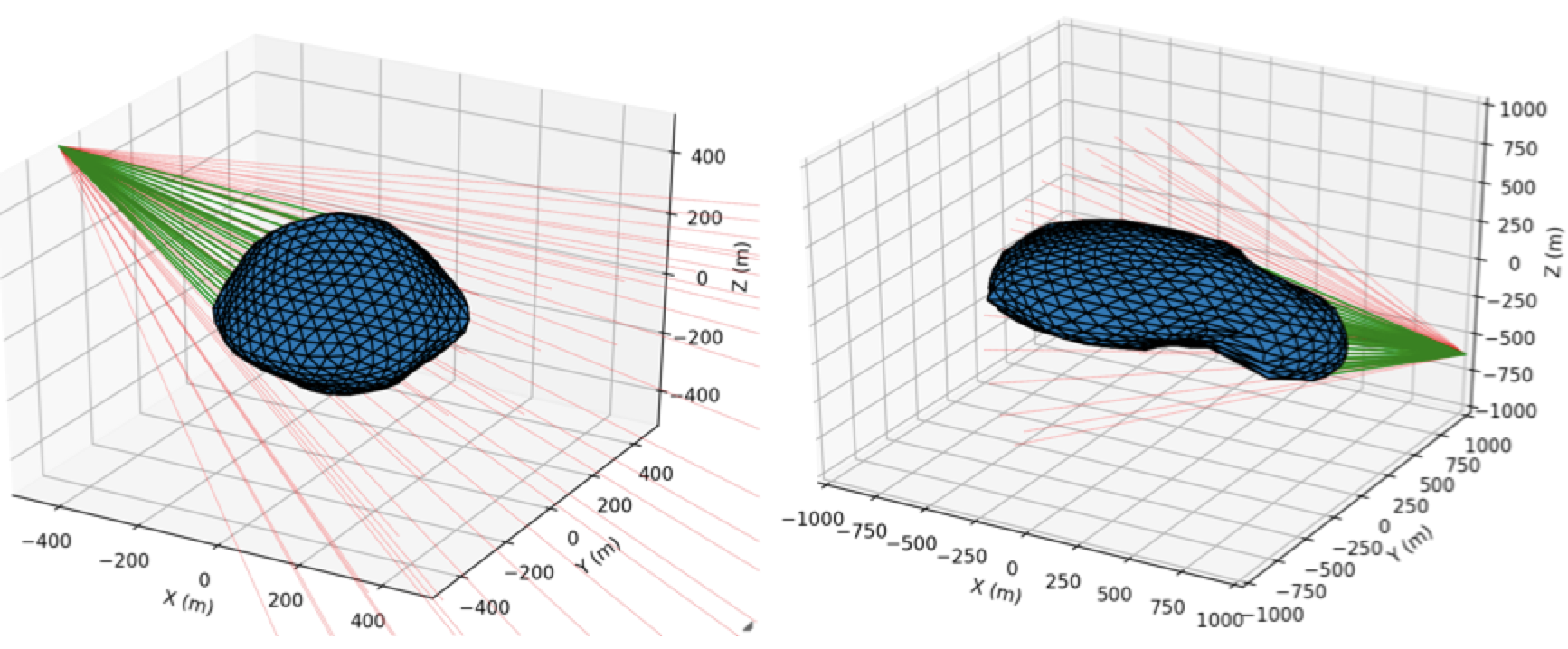}
\caption{Asteroids RQ36 (left) and Itokawa 3X (right)}
\label{fig:rq36_and_itokawa3x}
\end{center}
\end{figure}

\begin{table}[h!]
	\fontsize{10}{10}\selectfont
    \caption{Initial Condition altitudes for Real Shape Models}
   \label{tab:real_IC}
        \centering 
   \newcolumntype{R}{>{\raggedleft\arraybackslash}p{2.8cm}}
   \begin{tabular}{ l | R | R    } 
      \hline
      Asteroid & Min Altitude (m) & Max Altitude (m)  \\
      \hline
      rq36 & 100 & 500  \\
      rq36 EXT & 10 & 500 \\
      Itokawa & 100 & 250  \\
      Itokawa EXT & 10 & 250  \\
      Itokawa3X & 100 & 600  \\
      Itokawa3X EXT & 10 & 600  \\
   \end{tabular}
\end{table}

\begin{table}[h!]
	\fontsize{10}{10}\selectfont
    \caption{Performance with Real Shape Models}
   \label{tab:real_results}
        \centering 
   \newcolumntype{R}{>{\raggedleft\arraybackslash}p{2.5cm}}
   \begin{tabular}{ l | R | R | R   } 
      \hline
      Asteroid & Good Hover 1 (\%) & Good Hover 2 (\%) & Max Pos Err (m) \\
      \hline
      rq36 & 100.00 & 100.00 & 1.63\\
      rq36 EXT & 99.30 & 100.00 & 4.86 \\
      Itokawa & 99.84 & 99.95 & 8.37  \\
      Itokawa EXT & 99.80 & 99.98 &  9.32\\
      Itokawa3X & 99.94 & 100.00 & 4.41 \\
      Itokawa3X EXT & 98.02 & 99.64 & 17.52 \\
   \end{tabular}
\end{table}


\section{Conclusion}

We formulated a particularly difficult problem that to our knowledge has not been solved: precision hovering in an asteroid's body-fixed frame without a shape model or navigation aids, and without knowledge of the asteroid's environmental dynamics. To solve this problem we created a high fidelity 6-DOF simulator that synthesized asteroid models with shapes taking the form of asymmetric ellipsoids.  For purposes of computing angular velocity, the asteroid is modeled as a uniform density ellipsoid that in general is not rotating about a principal axis, resulting in time varying dynamics. We then optimized an adaptive policy that maps flash LIDAR sensor measurements directly to actuator commands.  The policy was optimized using reinforcement meta-learning, where the policy and value function networks each contained a recurrent hidden layer, and with the policy network using a convolutional front-end. During optimization, the agent was confronted with a new randomly generated asteroid for each episode, with randomized shape, density, rotational speed, and nutation angle. We then tested the policy, and demonstrated that the optimized policy generalizes well to novel hovering altitudes, hovering duration, actuator failure, actuator noise, sensor noise, and asteroid shapes and textures. Finally, we demonstrated hovering using an rq36 shape model and a scaled up Itokawa shape model. Comparing test performance for novel scenarios to that of scenarios used for optimization, we found similar, but slightly worse performance for the novel scenarios. Future work could improve upon this performance by creating synthetic asteroids with varying textures and more complex morphology, optimizing with random actuator failure and sensor noise, and exploring different convolutional network architectures. In addition, robustness to larger initial attitude errors and higher hovering altitudes would be enhanced by optimizing a separate policy that, prior to initiation of hovering, rotates the stabilized flash LIDAR seeker in a manner that minimizes the number of beams missing the asteroid. We expect that the ability to hover in the body-fixed frame immediately upon arrival at an asteroid will simplify shape model generation and other aspects of mission planning.

\bibliography{references}   

\end{document}